\documentclass[12pt]{article}
\usepackage{dina4p}
\usepackage{epsfig}

\newcommand{\beq}{\begin{equation}}
\newcommand{\eeq}{\end{equation}}
\newcommand{\beqn}{\begin{eqnarray}}
\newcommand{\eeqn}{\end{eqnarray}}

\begin{document}

\vspace*{-12mm}

\begin{flushright}
   DESY 01--109    \\
   August 2001     
\end{flushright}

\begin{center}  
\begin{Large} 
{\bf Future Measurements of Deeply Virtual \\ [2mm]
Compton Scattering at HERMES} 
\end{Large}

  \vspace*{4mm}

  \begin{large}
V. A. Korotkov$^{a,b}$ and W.-D. Nowak$^a$
  \end{large}

{$^a$ DESY Zeuthen, D-15735~Zeuthen, Germany \\
$^b$ IHEP, RU-142284 Protvino, Russia }
\end{center}

\begin{center}
{\bf Abstract}
\end{center}
Prospects for future measurements of Deeply Virtual Compton Scattering 
at HERMES are studied using different simple models for parameterizations of 
generalized parton distributions (GPDs). Measurements of the lepton charge and 
lepton beam helicity asymmetry will yield important input for theoretical
models towards the future extraction of GPDs.

\vspace*{-3mm}

\section{Introduction}

\vspace*{-2mm}

The study of hard exclusive processes in the Bjorken limit is now considered
as a promising tool to gain new insight into details of the nucleon structure 
that cannot be studied with inclusive deep inelastic scattering (DIS). 
A unified theoretical
description of hard exclusive and inclusive processes has been obtained
through the formalism of generalized parton distributions
\cite{muller94,Ji,rad} (see recent reviews in Ref.s \cite{radreview,gpv01})
which are also called skewed, off-forward or non-forward parton distributions.

An ordinary parton distribution represents the probability to find
a parton with a specified longitudinal momentum fraction $x$ in the
fast moving hadron and thus is summing over all partonic configurations
containing such a parton. In contrast, GPDs represent the interference
of two different wave functions, one with a parton having momentum fraction
$x + \xi$ and another one with a parton having momentum fraction  $x - \xi$.
GPDs, besides the longitudinal momentum fraction variables $x$ and $\xi$ 
(called skewedness parameter), depend on a third independent variable, 
the momentum transfer $\Delta^2 = (p - p')^2$ between initial and final nucleon
states with momenta $p$ and $p'$, respectively.

There are four different types of quark GPDs contributing
to the simplest hard exclusive process: Deeply Virtual Compton Scattering
(DVCS), $e p \longrightarrow e p \gamma$. In the unpolarized distributions,
$H^q(x,\xi,\Delta^2)$ and $E^q(x,\xi,\Delta^2)$, the quark helicities are
summed over. The polarized distributions, $\widetilde{H}^q(x,\xi,\Delta^2)$ 
and $\widetilde{E}^q(x,\xi,\Delta^2)$, are responsible for the differences 
between right- and left-handed quarks.

The generalized parton distributions combine the characters of both the 
ordinary parton distributions and of nucleon form factors. On the one hand, 
in the limit $\Delta^2 \rightarrow 0$, $\xi \rightarrow 0$ holds
\begin{equation}
\label{eq:spdtoq}
     H^q(x,0,0) = q(x),~~~ \widetilde{H}^q(x,0,0) = \Delta q(x),
\end{equation}
where $q(x)$ and $\Delta q(x)$ are the ordinary quark number density and quark 
helicity distributions. On the other hand, the first moment of GPDs 
must satisfy the following sum rules, 
\newpage
\beqn
\label{eq:spdtoff}
     \int^1_{-1} dx H^q(x,\xi,\Delta^2) &=& F^q_1(\Delta^2) \ ,  \nonumber \\
     \int^1_{-1} dx E^q(x,\xi,\Delta^2) &=& F^q_2(\Delta^2) \ ,  \nonumber \\
     \int^1_{-1} dx \widetilde{H}^q(x,\xi,\Delta^2) &=& g^q_A(\Delta^2) \ , 
           \nonumber \\ 
     \int^1_{-1} dx \widetilde{E}^q(x,\xi,\Delta^2) &=& h^q_A(\Delta^2) \ ,
\eeqn
where $F_1^q(\Delta^2)$ and $F_2^q(\Delta^2)$ are the Dirac and Pauli form 
factors and $g_A^q(\Delta^2)$ and $h_A^q(\Delta^2)$ are the axial-vector and 
pseudo-scalar form factors, respectively.

In the above formulae the variable $x$ is defined in the range
$(-1, +1)$ and negative values of it correspond to anti-quark
distributions in the following manner:
\beq
 q(-x) = -\bar{q}(x), \, \, \, \Delta q(-x) = \Delta \bar{q}(x) .
\eeq
Two different regions exist for GPDs with respect to the variables $x$ and
$\xi$. For $|x| > \xi$ the GPDs are the generalizations of the ordinary parton
distributions, while for $|x| < \xi$ the GPDs behave like meson
distribution amplitudes.

The recent strong interest in GPDs was stimulated by the finding of 
Ji \cite{Ji} that the second moment of the unpolarized GPDs
at $\Delta^2 = 0$ is relevant to the spin structure of the nucleon since it
determines the total quark angular momentum:
\begin{equation}
J_{q}\, =\, {1\over 2}\, \int _{-1}^{+1}dx\, x\, 
\left[ H^{q}(x,\xi ,\Delta^2=0)+E^{q}( x,\xi ,\Delta^2=0)\right] \; .
\label{eq:dvcsspin}
\end{equation}
The total quark angular momentum \( J_{q} \) decomposes as 
\vspace*{-1ex}
\begin{equation}
\label{eq:spindecomp}
J_{q}={1\over 2}\Delta \Sigma +L_{q}\; ,
\end{equation}
 where \( \Delta \Sigma /2 \) and \( L_{q} \) denote quark spin
and orbital angular momentum, respectively. 
As \( \Delta \Sigma  \) is measured through polarized DIS experiments, 
a measurement of $J_{q}$ through the sum rule Eq. (\ref{eq:dvcsspin}) in
terms of GPDs provides a model-independent way to determine the contribution 
of the quark orbital momentum to the nucleon spin. Eventually even
the contribution of the total gluon angular momentum \( J_{g} \) may become 
accessible through
\vspace*{-1ex}
\beq
{1\over 2}\, =\, J_{q}\, +\, J_{g}\; .
\eeq
\vspace*{-3ex}

First measurements of the DVCS lepton helicity asymmetry have been
accomplished recently by HERMES \cite{hermesdvcs} at 27.5 GeV and by 
CLAS \cite{CLASdvcs} at 4.25 GeV. Several plans exist for further
measurements of DVCS and of other hard exclusive reactions to accomplish
a first insight into GPDs. At HERMES, an upgrade of the spectrometer with 
a recoil detector \cite{prcrecoil} will significantly improve the separation 
of exclusive events.
The DVCS process has also been observed in $e^+ p$ collider experiments
at DESY by ZEUS \cite{desyzeus} and H1 \cite{desyh1}. The DVCS cross section
was measured and compared to the QCD-based predictions. Measurements of the
lepton beam helicity asymmetry by H1 and ZEUS will become possible in the 
near future when longitudinally polarized leptons will be made available also
to the collider experiments at HERA.

The main aim of this paper is the evaluation of the anticipated
statistical accuracy for future measurements of DVCS asymmetries at 
HERMES upgraded with the recoil detector. In section 2 of the paper different 
versions of GPD parameterizations are discussed. The third section deals with
the assessment of the expected size of the asymmetries, based upon different
GPD parameterizations, and with the evaluation of their projected statistical 
accuracy. Finally, the conclusions of the paper are presented.

\vspace*{-2ex}

\section{Parameterization of Generalized Parton Distributions}

\vspace*{-1ex}

Two examples of GPD calculations are presently known in the literature.
While bag model calculations \cite{jisong} show a weak dependence
of the distributions on the skewedness parameter, chiral quark soliton 
model calculations \cite{petrov98}, in contrast, show a strong
dependence on $\xi$. The common approach at the moment is to use a guess
that satisfies general constraints on GPDs known
from theory. This paper basically follows the ansatz
proposed in Ref.s \cite{vgg99,gpv01}. Here, the dependence of the GPDs
on $\Delta^2$ is taken in a factorized form with respect to the other 
variables whereby satisfying Eq.(\ref{eq:spdtoff}). 
Any scale dependence of the GPDs is neglected.

In the simplest approach GPDs can be assumed to be independent of the
skewedness parameter $\xi$.
In the following, only $u$- and $d$-quark GPDs are considered to be non-zero. 
The function $H$, for example, is written as a product of an ordinary quark 
distribution function and a form factor
\vspace*{-1.5ex}
\beqn
H^u(x, \xi, \Delta^2) \,&=&\, u( x ) \; F_1^u(\Delta^2) \, / \, 2\;, \nonumber\\
H^d(x, \xi, \Delta^2) \,&=&\, d( x ) \; F_1^d(\Delta^2) \; .
\label{eq:hnoskew}
\eeqn
Here \( u(x) \) and \( d(x) \) are the unpolarized quark distributions 
and $F_1^{u(d)}(\Delta^2)$ are defined through the electro-magnetic
form factors of proton and neutron:
\beq
 F_1^u = 2 F_1^p + F_1^n, \, \, F_1^d = F_1^p + 2 F_1^n .
\label{eq:fpntofud}
\eeq
In the same context, the function $\widetilde{H}$ is written to be related to 
quark helicity distributions and axial-vector form factors:
\vspace*{-1.5ex}
\beqn
\widetilde{H}^u (x, \xi, \Delta^2) \,&=&\, \Delta u_V( x ) 
\;{g_{A}^{u}(\Delta^2)}/{g_{A}^{u}(0)}\;, \nonumber \\
\widetilde{H}^d (x, \xi, \Delta^2) \,&=&\, \Delta d_V( x ) 
\;{g_{A}^{d}(\Delta^2)}/{g_{A}^{d}(0)} \; , 
\label{eq:htnoskew}
\eeqn
where $g_A^u = \frac{1}{2}g_A + \frac{1}{2}g_A^0$,
$g_A^d = -\frac{1}{2}g_A + \frac{1}{2}g_A^0$ and $g_A^0 = \frac{3}{5}g_A$. 

The functions $E$ and $\widetilde{E}$ have no definite limit at
$\Delta^2 \rightarrow 0$,
$\xi \rightarrow 0$, as it exists for the functions $H$ and $\widetilde{H}$ 
(cf. Eq. (\ref{eq:spdtoq})). 
In absence of any other guide the ansatz for $E$ is chosen in a form 
analogous to the function $H$:
\vspace*{-1ex}
\beqn
E^u(x, \xi, \Delta^2) \,&=&\, u( x ) \; F_2^u(\Delta^2) \, / \, 2\;, \nonumber\\
E^d(x, \xi, \Delta^2) \,&=&\, d( x ) \; F_2^d(\Delta^2) \; ,
\label{eq:enoskew}
\eeqn
where $F_2^{u(d)}$ is defined in the same way as $F_1^{u(d)}$ in
Eq. (\ref{eq:fpntofud}).

The function $\widetilde E$ is modelled to be due to the pion pole:
\vspace*{-1ex}
\beq
\widetilde E^{u}(x, \xi, \Delta^2) = -\widetilde E^{d}(x, \xi, \Delta^2) = 
{1\over 2}\; \widetilde E_{\pi -pole} (x, \xi, \Delta^2) \; ,
\label{eq:etpole}
\eeq
\beq
\widetilde E_{\pi -pole}(x, \xi, \Delta^2) \; =\; 
\theta \left( -\xi \leq x\leq \xi \right) 
\; h_{A}(\Delta^2)\; {1\over \xi }\; \Phi \left( {x\over \xi }\right) \; ,
\label{eq:pipole}
\eeq
\newpage
where $ \Phi (z) = 3/4  (1-z^{2}) $ is the pion distribution amplitude,
$h_A (\Delta^2 ) = {{4 M^2 g_A}\over{m_\pi^2 - \Delta^2}}$, and 
$\theta ( x )$ is the usual step function. 
 
To introduce, as a next step, a dependence of the GPDs on the skewedness 
parameter $\xi$,
the double-distribution formalism \cite{muller94,rad} can be used. 
In this model, the $\Delta^2$-independent 
part of the function $H$ can be written in the following form
\beq
H^q_{DD}(x,\xi) = \int_{-1}^{1}dy\ \int_{-1+|y|}^{1-|y|} dt\
\delta(x - y - t\xi)\  h( y, t)\ q(y) \ \, .
\label{eq:doubled}
\eeq
Here $q(y)$ is the ordinary quark distribution and $h( y, t)$ is the so-called
profile function:
\beq
h( y, t) = 
 \frac{\Gamma(2b+2)}{2^{2b+1}\Gamma^2(b+1)}\
\frac{\bigl[(1-|y|)^2-t^2\bigr]^{b}}{(1-|y|)^{2b+1}}\; ,
\label{eq:proffun}
\eeq
with $b$ as a free parameter and $b \rightarrow \infty$ corresponding to the 
skewedness-independent parameterization. Analogous expressions can 
be written for the functions $\widetilde H$ and $E$.

The functions $H$ and $E$ in the form of double distributions lack the 
correct polynomiality properties of GPDs. However, they can be restored 
by introducing the so-called $D$-term \cite{polweiss}. The $D$-term 
contribution has different sign for $H$ and $E$ and hence cancels 
in Ji's angular momentum sum rule Eq. (\ref{eq:dvcsspin}). The full
model expressions for the GPDs therefore have the following form:
\vspace*{-2ex}
\beqn
H^q(x, \xi) \,&=&\, H^q_{DD}(x, \xi) \,+\,
\theta(\xi-|x|)\, \frac{1}{N_f} \, D(\frac{x}{\xi})\, , \nonumber \\
E^q(x, \xi) \,&=&\, E^q_{DD}(x, \xi) \,-\,
\theta(\xi-|x|)\, \frac{1}{N_f} \, D(\frac{x}{\xi})\, ,
\label{eq:hefull}
\eeqn
where $N_f$ = 3 is the number of active flavors. The parameterization
of the $D$-term is taken in a form that follows from chiral
quark soliton model calculations \cite{kivpolvan}. 
For more details on possible GPD parameterizations and their
discussion we refer to Ref.s \cite{vgg99, gpv01}.

Quark number density and quark helicity distributions $q(x)$ and
$\Delta q(x)$, respectively, were used in the parameterizations of Ref. 
\cite{mrst98} (at $Q^2 = 2$~GeV$^2$) and Ref. \cite{leader} 
(at $Q^2 = 1$~GeV$^2$).

In the projections described below five different versions of GPD
parameterizations were included: \vspace*{-1ex}
\begin{itemize}
 \item [(A)] $H( x, \xi)$, $\widetilde{H}( x, \xi)$, and $E( x, \xi)$ are 
skewedness independent and given 
by Eq.s (\ref{eq:hnoskew}), (\ref{eq:htnoskew}), and (\ref{eq:enoskew}),
respectively.  \vspace*{-1ex}
 \item [(B)] The skewedness dependence of $H( x, \xi)$, 
$\widetilde{H}( x, \xi)$, 
and $E( x, \xi)$ is generated 
through the double-distribution formalism Eq.s (\ref{eq:doubled}) and 
(\ref{eq:proffun}) with the parameter $b = 1$. \vspace*{-1ex}
 \item [(C)] same as (B), but with the parameter $b = 3$. \vspace*{-1ex}
 \item [(D)] same as (B), but the contribution of the $D$-term to $H( x, \xi)$ 
and $E( x, \xi)$ is included additionally, according to Eq.(\ref{eq:hefull}). 
\vspace*{-1ex}
 \item [(E)] same as (D), but with the parameter $b = 3$. \vspace*{-1ex}
\end{itemize}
For $\widetilde{E}( x, \xi)$ the pion pole expression Eq. (\ref{eq:etpole})
is used for all versions of GPDs.

\newpage

\begin{figure}[ht]
\centering
\epsfig{file=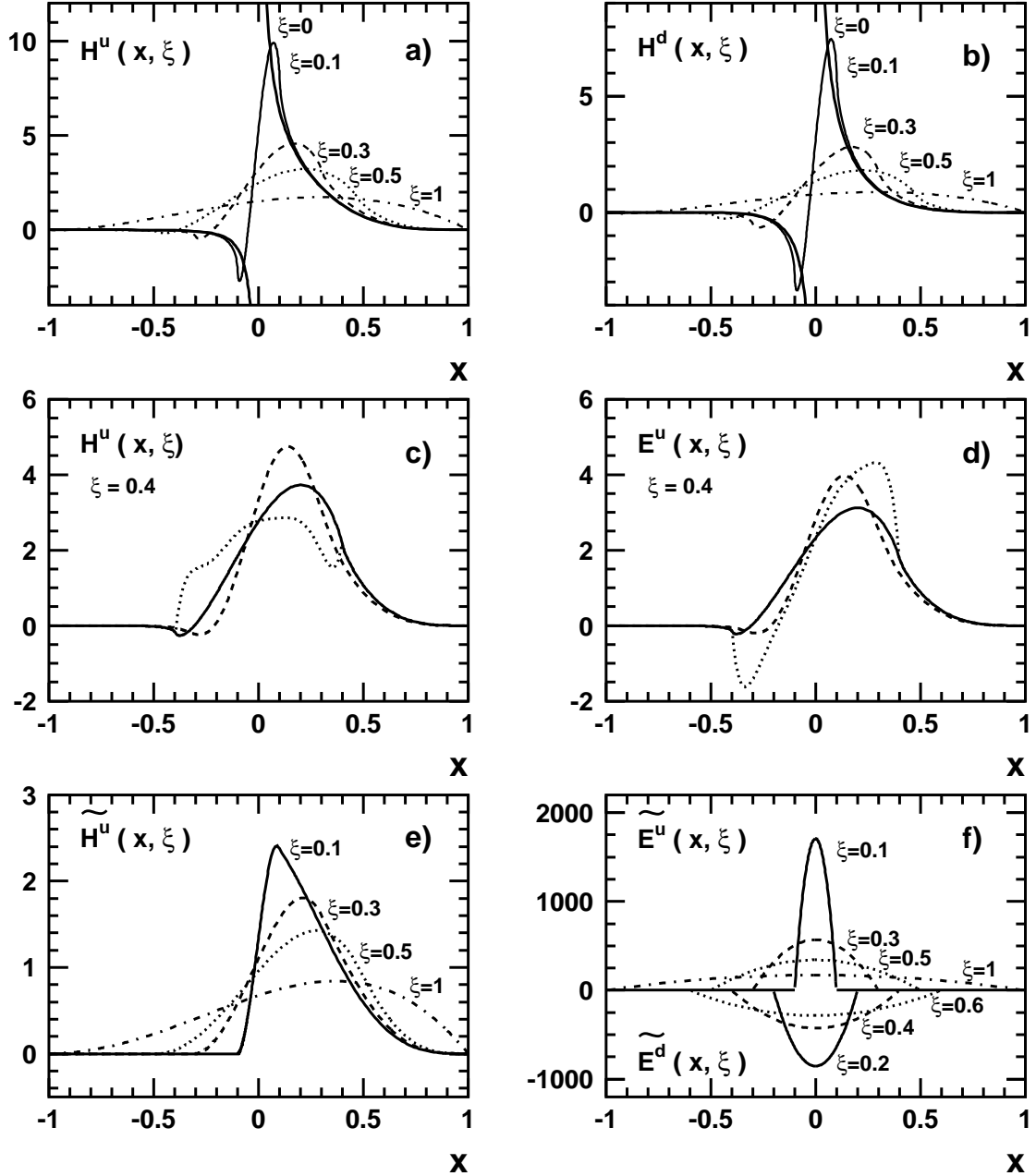,width=18.0cm}
\vspace*{-1cm}
\caption{\it Illustration of the $x$-behaviour of various generalized parton 
distributions for different values of the skewedness parameter $\xi$
(at $\Delta^2 = 0$). For explanations see text. }
\label{fig:allspd}
\end{figure}

The dependences of the GPDs on the variables $x$ and $\xi$ (at $\Delta^2 = 0$),
obtained from these five different parameterizations, are displayed for a
few examples in fig. \ref{fig:allspd}.
Panels a) and b) show the GPD $H$ in version (B) for different values of $\xi$,
compared between $u$- and $d$-quark. Panels c) and d) show the unpolarized 
$u$-quark GPDs $H$ and $E$ at fixed $\xi = 0.4$, compared between different 
versions: (B) as solid line, (C) as dashed line and (D) as dotted line.
Panels e) and f) show the polarized $u$-quark GPDs $\widetilde{H}$ and 
$\widetilde{E}$ in version (B) compared for different values of $\xi$. 
Although $\widetilde{E}$ does not appear in the DVCS asymmetries (see below), 
its $u$- vs. $d$-quark comparison is shown, as well. By definition 
$\widetilde{E}$ is large at $\Delta^2 = 0$ (see Eq.~(\ref{eq:etpole})).

\section{Deeply Virtual Compton Scattering}

\begin{figure}[hbt]
\centering 
\begin{minipage}[c]{6.0cm}
\raggedleft
\epsfig{file=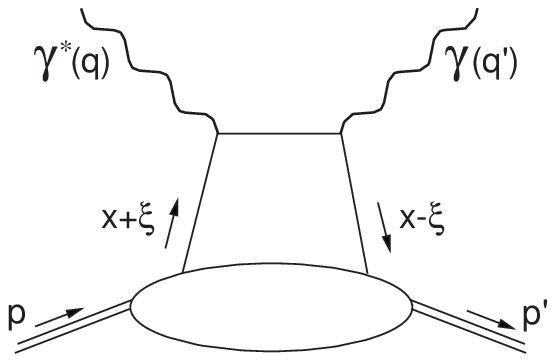,width=6.0cm}
\vspace*{-1cm}
\caption{\it Leading handbag 
diagram for DVCS. }
\label{fig:handbag}
\end{minipage}
\hspace*{1.0cm}
\begin{minipage}[c]{9.0cm}
\raggedright
\epsfig{file=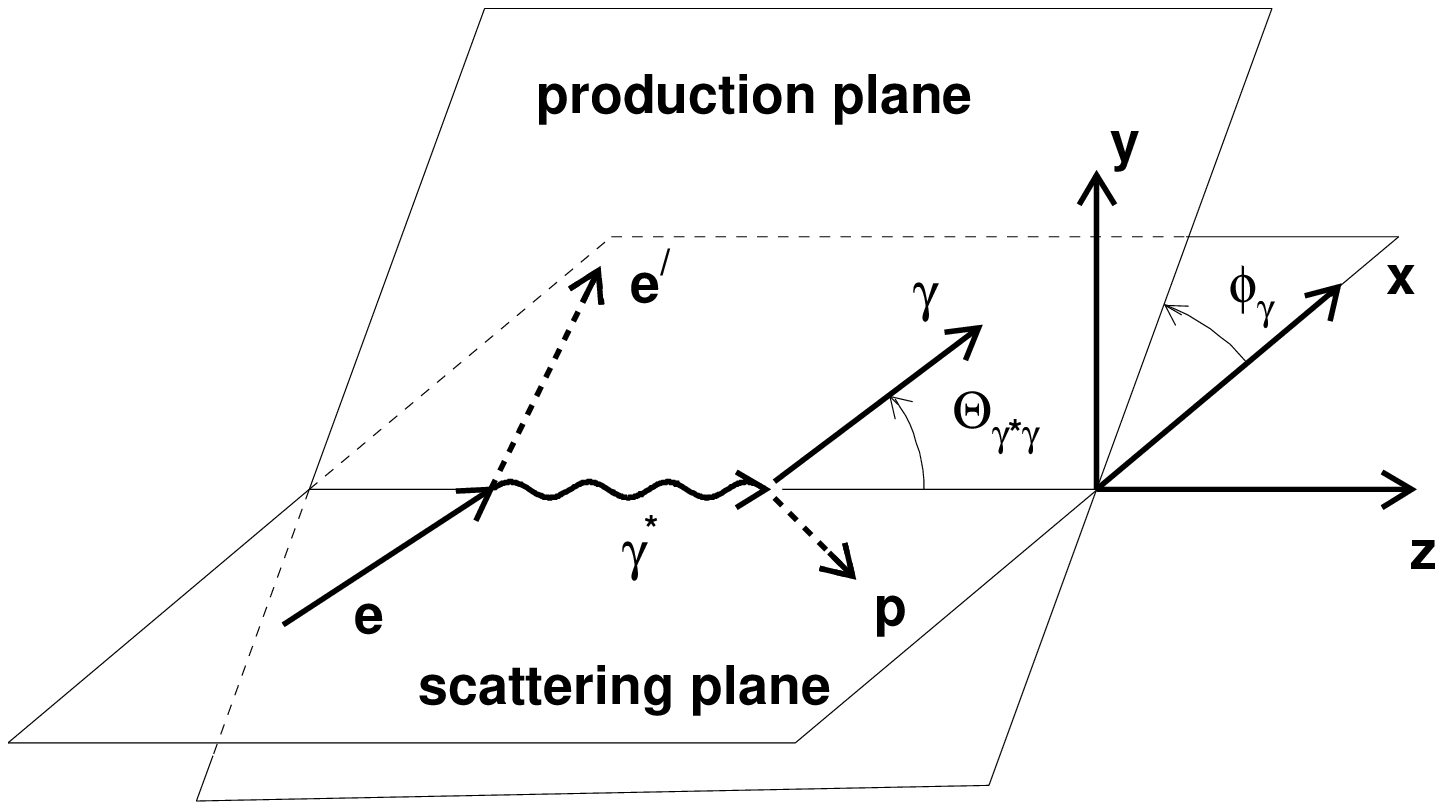,width=9.0cm}
\caption{\it Kinematic configuration for the process
$e p \longrightarrow e p \gamma$. }
\label{fig:gplane}
\end{minipage}
\end{figure}

In the Bjorken limit, DVCS is dominated by the handbag diagram 
(fig. \ref{fig:handbag}) and its amplitude can be factorized 
into a soft part described by GPDs and a hard part representing
a parton process calculable in perturbative QCD. In this limit 
the skewedness parameter $\xi$ can be related to the Bjorken variable $x_B$:
\begin{equation}
\xi = \frac{x_B / 2}{1 - x_B/2}
\end{equation}
The same final state, $e p \gamma$,
can also be produced via the Bethe-Heitler (BH) process in which an electron
scatters elastically off the target proton and the initial or 
final state electron radiates a real photon. The cross section of this 
process can be calculated exactly once the Dirac and Pauli nucleon 
form-factors are known. On the one hand, the BH process 
constitutes the main background to DVCS, on the other hand their interference
opens the unique opportunity for independent measurements of the real and 
imaginary parts of a certain DVCS amplitude combination (see below).

\begin{figure}[hbt]
\centering
\epsfig{file=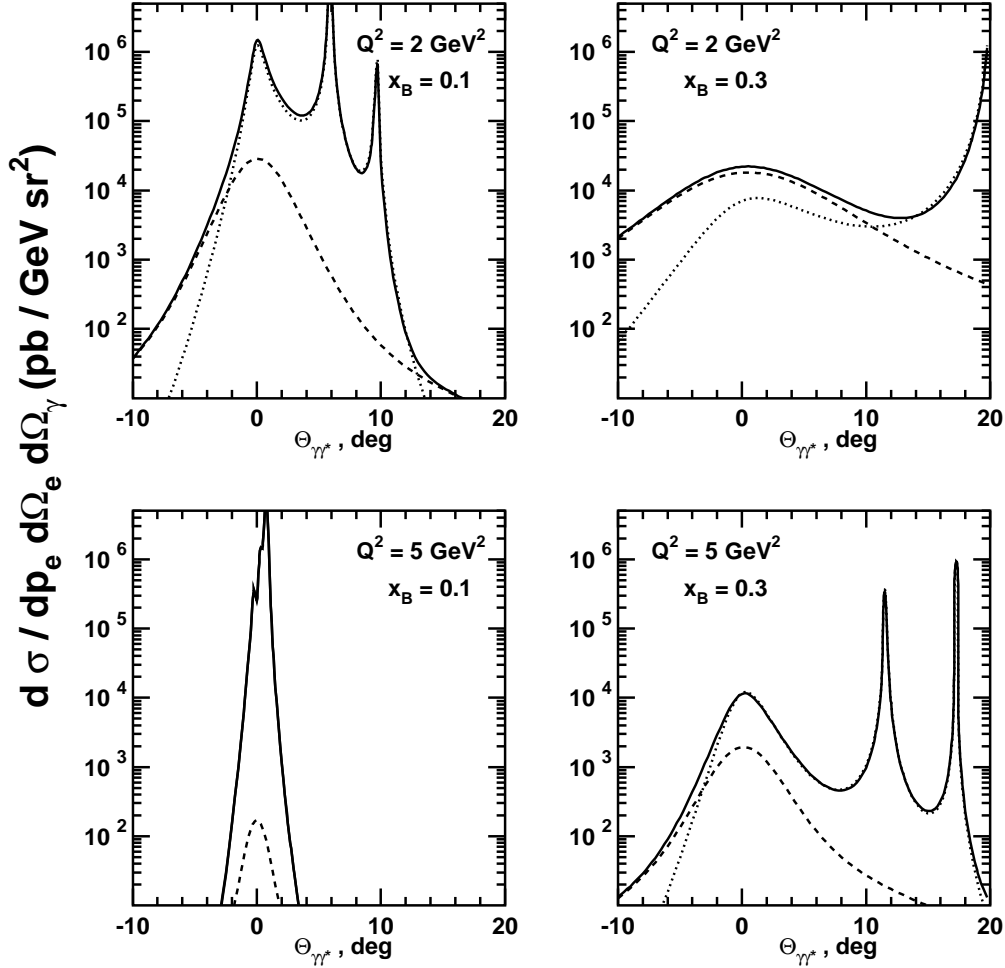,width=15.0cm}
\caption{\it Illustration of the behaviour of the differential in-plane 
cross-section as a function of the polar angle between the virtual and 
the real photon for DVCS (dashed lines),
Bethe-Heitler (dotted lines) and total $\gamma$ production (solid lines)
in $e^+ p$ interactions at HERMES energy $E_e = 27.5$~GeV. Different panels
are for different values of $x_B$ and $Q^2$. }
\label{fig:xsect}
\end{figure}

The cross-section of the DVCS process and its interference with the BH
process has been considered in a number of papers \cite{Ji,rad}.
The kinematic configuration of the process $e p \longrightarrow e p \gamma$
is shown in fig. \ref{fig:gplane}. Here $\phi_\gamma$ describes the azimuthal 
orientation of the production plane (comprising $\gamma^*$, $\gamma$ and $p$)
relative to the scattering plane (comprising initial and final lepton, as
well as the virtual photon). The laboratory polar angle between virtual
and real photon is denoted by $\theta_{\gamma \gamma^*}$.
In-plane (i.e. $\phi_\gamma = 0$)
differential cross-sections for DVCS, BH and
total $\gamma$ production in $e^+ p$ interactions at the HERMES energy of
$E_e = 27.5$~GeV were calculated following Ref. \cite{Ji} and are
presented in fig. \ref{fig:xsect}. The DVCS cross-section shows a maximum at
$\Theta_{\gamma \gamma^*} = 0$, while the BH cross-section
has a three-pole structure due to the propagators of the virtual electrons
and the virtual photon. It is apparent that at HERMES energies the 
cross-section of DVCS is smaller than that of BH over practically the entire 
kinematic region. Therefore measurements of GPDs at HERMES must be based upon
the interference between the DVCS and BH processes.

In Ref. \cite{belitsky} amplitudes of DVCS, BH and of the interference 
terms were calculated at the leading twist-2 level for 
polarized and unpolarized initial particles. This approach
was used to determine the below presented projections
for future measurements of DVCS-BH interference effects at HERMES. 

The primary interest of HERMES in the context of the planned recoil detector
upgrade lies with measurements using an unpolarized proton target.
In this case two different types of experiments are possible that will be 
giving insight into GPDs:

I) Measurements of the lepton charge asymmetry with unpolarized leptons
of either charge:
\begin{equation}
\label{eq:lchasy}
d\Delta \sigma_{ch}
\equiv d \sigma(e^{+} p) - d \sigma(e^{-} p) \sim
\cos(\phi_\gamma) \times 
{\rm Re} \left\{F_1 {\cal H}_1
+ \frac{x_B}{2 - x_B} (F_1+F_2) \widetilde{\cal H}_1
- \frac{\Delta^2}{4 M^2} F_2 {\cal E}_1\right\} 
\end{equation}
This asymmetry allows access to the real parts of the DVCS amplitudes
${\cal H}_1$, $\widetilde{\cal H}_1$, and ${\cal E}_1$. 

II) Measurements of the beam helicity asymmetry using a polarized positron 
beam:
\begin{equation}
\label{eq:bhelasy}
d\Delta \sigma_{LU}
\equiv d\sigma(\overrightarrow{e^+}p) - 
         d\sigma(\overleftarrow{e^+}p) \sim  
\sin(\phi_\gamma) \times
{\rm Im} \Bigg\{ F_1 {\cal H}_1
+ \frac{x_B}{2 - x_B} (F_1 + F_2) \widetilde{\cal H}_1
- \frac{\Delta^2}{4 M^2} F_2 {\cal E}_1 \Bigg\}
\end{equation}
This asymmetry allows access to the imaginary parts of the same
amplitudes.

The imaginary and real parts of the DVCS amplitudes ${\cal H}_1$ and 
$\widetilde{\cal H}_1$ are related to the GPDs $H$ and $\widetilde{H}$
as follows:  
\begin{eqnarray}
 {\rm Im } \, {\cal H}_1 &=& 
- \pi \sum_q e^2_q \bigl( H( \xi, \xi, \Delta^2) - 
                          H( -\xi, \xi, \Delta^2) \bigr)\, , \nonumber \\
 {\rm Im } \, \widetilde{\cal H}_1 &=& 
- \pi \sum_q e^2_q \bigl( \widetilde{H}( \xi, \xi, \Delta^2) + 
                          \widetilde{H}( -\xi, \xi, \Delta^2) \bigr) \, ,
\nonumber \\
 {\rm Re } \, {\cal H}_1 &=& 
\sum_q e^2_q \Bigl[ P \int_{-1}^{+1} H( x, \xi, \Delta^2) \,
\Bigl( \frac{1}{x - \xi} + \frac{1}{x + \xi} \Bigr) dx \, \Bigr] \, ,
\nonumber \\
 {\rm Re } \, \widetilde{\cal H}_1 &=&
\sum_q e^2_q \Bigl[ P \int_{-1}^{+1} \widetilde{H}( x, \xi, \Delta^2) \,
\Bigl( \frac{1}{x - \xi} - \frac{1}{x + \xi} \Bigr) dx \, \Bigr] \, ,
\end{eqnarray}
where $P$ denotes Cauchy's principal value.
The DVCS amplitudes ${\cal E}_1$ and $\widetilde{\cal E}_1$ can be expressed 
through $E$ and $\widetilde E$ analogously.

Projections for statistical accuracies attainable in measurements of the 
lepton charge and lepton beam helicity asymmetry at HERMES were calculated 
for an integrated luminosity of $2$~fb$^{-1}$ \cite{prcrecoil} which 
corresponds to the expected value for one year of data taking. 
The HERMES geometrical acceptance for the detection of the scattered 
electron, the photon and the recoil proton was taken into account.
The following kinematic cuts\footnote{Here $E_e$ and $E_\gamma$ are the
energy of the incoming electron and the outgoing photon, respectively, while
$P_p$ is the momentum of the outgoing proton. The standard DIS variables
$Q^2$ and $W^2$ describe the four-momentum transfer and the energy of the 
$\gamma^* p$ system, respectively. The laboratory polar angle of the outgoing 
proton is given by $\theta_p^{lab}$.} were applied:
$E_e > 3.5$~GeV, $W^2 > 4$~GeV$^2$, $Q^2 > 1$~GeV$^2$, 
$E_\gamma > 1$~GeV, $P_p > 0.2$~GeV, $0.35 < \theta_p^{lab} < 1.35$~rad,
and $15 < \Theta_{\gamma \gamma^*} < 70$~mrad.

To calculate the asymmetries Eq.s (\ref{eq:lchasy}) and (\ref{eq:bhelasy}) 
the 5-fold differential cross-section
has to be integrated over the appropriate kinematic region accounting for
the HERMES geometrical acceptance:
$$ \frac{d \sigma}{d \phi_\gamma} = 
\int \frac{d^5 \sigma}{dx dQ^2 d|\Delta^2| d\phi_\gamma d\phi_{el}} \,
dx dQ^2 d|\Delta^2| d\phi_\gamma d\phi_{el} \ , $$
where $\phi_{el}$ is the azimuthal angle of the scattered electron.
It is appropriate to define differences and sums of certain cross-sections:
\beqn
 \frac{d \Delta \sigma_{ch}}{d \phi_\gamma} \, &=& \,
\frac{d \sigma (e^+ p)}{d \phi_\gamma} - 
\frac{d \sigma (e^- p)}{d \phi_\gamma} ,   \nonumber \\
 \frac{d \Sigma \sigma_{ch}}{d \phi_\gamma} \, &=& \,
\frac{d \sigma (e^+ p)}{d \phi_\gamma} + 
 \frac{d \sigma (e^- p)}{d \phi_\gamma} ,   \nonumber \\
 \frac{d \Delta \sigma_{LU}}{d \phi_\gamma} \, &=& \,
\frac{d \sigma (e^\uparrow p)}{d \phi_\gamma} - 
\frac{d \sigma (e^\downarrow p)}{d \phi_\gamma} ,  \nonumber \\
 \frac{d \Sigma \sigma_{LU}}{d \phi_\gamma} \, &=& \,
\frac{d \sigma (e^\uparrow p)}{d \phi_\gamma} + 
 \frac{d \sigma (e^\downarrow p)}{d \phi_\gamma} .
\eeqn
Using these definitions the $\phi$-dependence of the the lepton charge 
asymmetry reads
\beq
A_{ch}( \phi ) = { {{\int_{\phi - \Delta\phi}^{\phi + \Delta\phi} d\phi \, 
                     d \Delta \sigma_{ch}/ d\phi } }  \over
          { \int_{\phi - \Delta\phi}^{\phi + \Delta\phi} d\phi \, 
             d \Sigma \sigma_{ch}/ d\phi }} \, ,
\label{eq:asychphi}
\eeq
while an integrated lepton charge asymmetry can be defined by forming the 
difference between two integrals over appropriately defined `halves' of the 
$\phi$-distribution:
\beq
\widetilde{A}_{ch} = 
    { {{\int_{-\pi/2}^{\pi/2}  d\phi \, d \Delta \sigma_{ch}/ d\phi} -
          {\int_{\pi/2}^{3 \pi/2} d\phi \, d \Delta \sigma_{ch}/ d\phi } }  
         \over
          { \int_0^{2 \pi} d\phi \, d \Sigma \sigma_{ch}/ d\phi }} \, .
\label{eq:asychx}
\eeq
Analogously the $\phi$-dependent and an integrated lepton beam helicity 
asymmetry, respectively, are defined as follows:
\beq
A_{LU}(\phi) = { {{\int_{\phi - \Delta\phi}^{\phi + \Delta\phi} d\phi \, 
                 d \Delta \sigma_{LU}/ d\phi} }  \over
          { \int_{\phi - \Delta\phi}^{\phi + \Delta\phi} d\phi \, 
                 d \Sigma \sigma_{LU}/ d\phi } } \, ,
\label{eq:asybhphi}
\eeq
\beq
\widetilde{A}_{LU} = { {{\int_0^\pi d\phi \, d \Delta \sigma_{LU}/ d\phi} -
          {\int_\pi^{2 \pi} d\phi \, d \Delta \sigma_{LU}/ d\phi } }  \over
          { \int_0^{2 \pi} d\phi \, d \Sigma \sigma_{LU}/ d\phi }} \, .
\label{eq:asybhx}
\eeq

\begin{figure}[ht]
\centering
\begin{minipage}[c]{8.0cm}
\centering
\epsfig{file=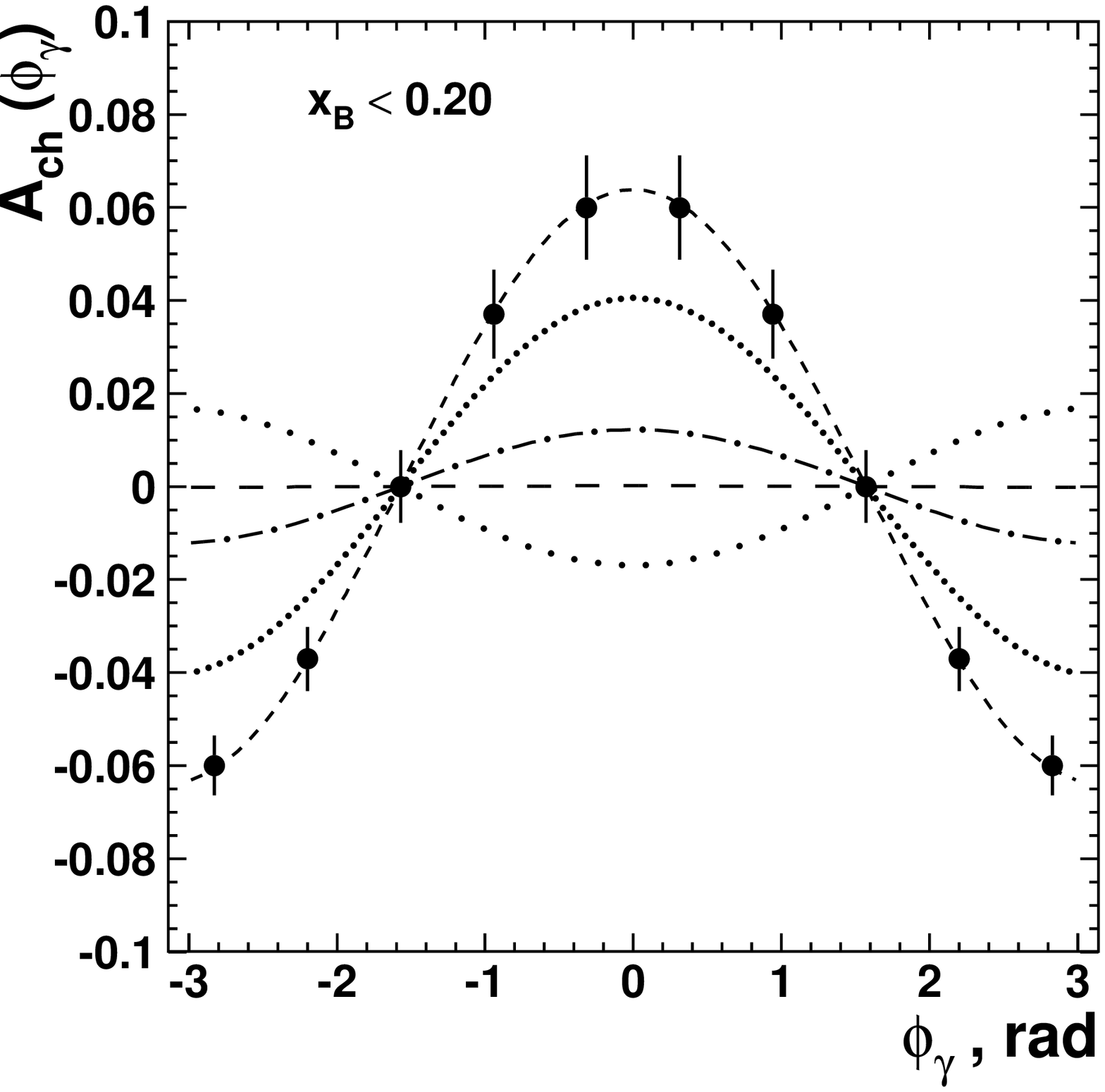,width=7.5cm}
\end{minipage}
\begin{minipage}[c]{8.0cm}
\centering
\epsfig{file=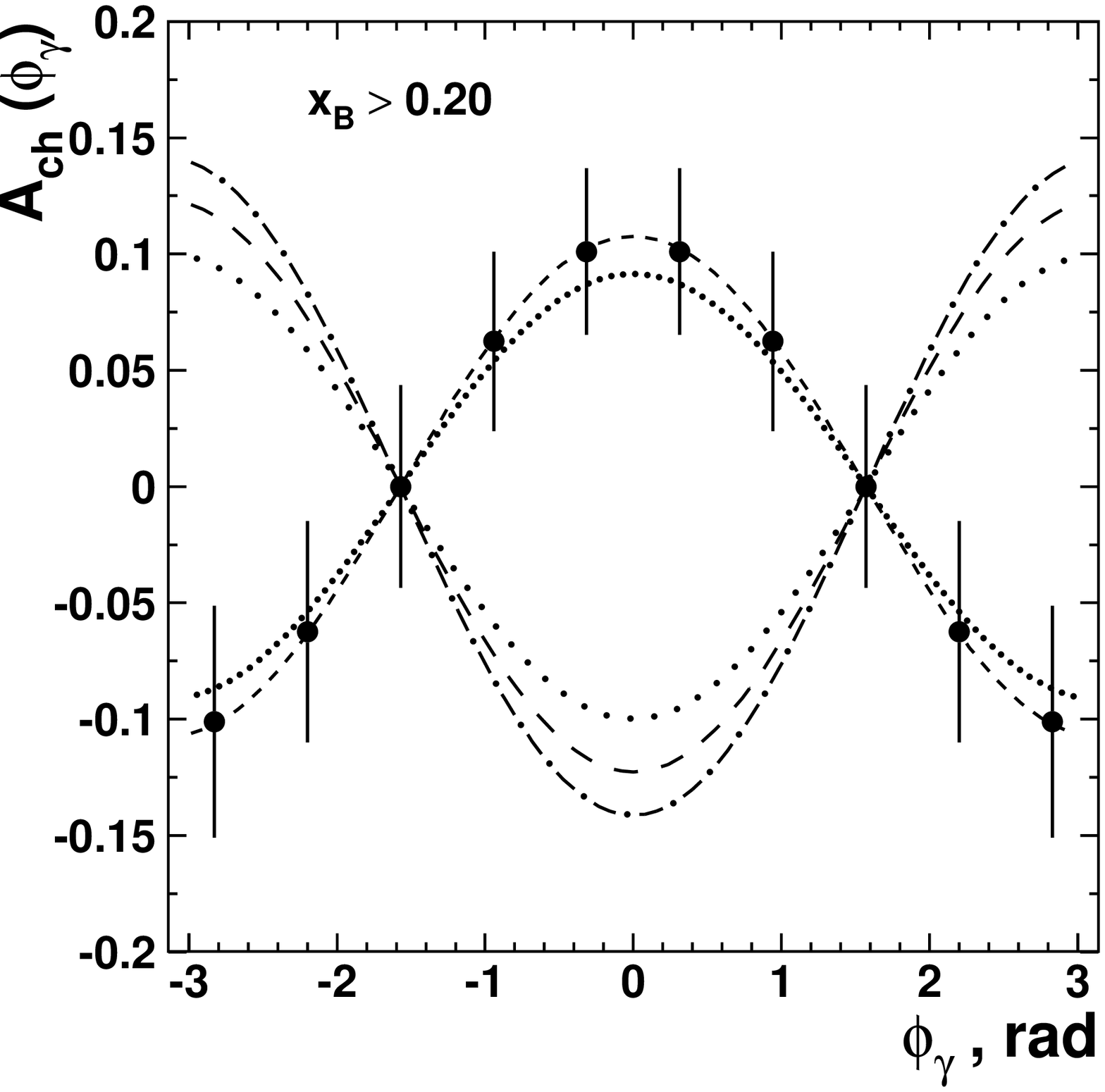,width=7.5cm}
\end{minipage}
\caption{\it Projected statistical accuracy for the
lepton charge asymmetry Eq. {\rm(\ref{eq:asychphi})} as a function of the 
azimuthal angle $\phi_\gamma$ between scattering plane and production plane. 
Predictions of different GPDs models (see text) are shown in two regions of 
$x_B$. Version (A) - dash-dotted line;
version (B) - long-space dotted line; version (C) - long-space dashed line;
version (D) - dotted line; version (E) - dashed line. }
\label{fig:lchphidep}
\end{figure}

\begin{figure}[ht]
\centering
\epsfig{file=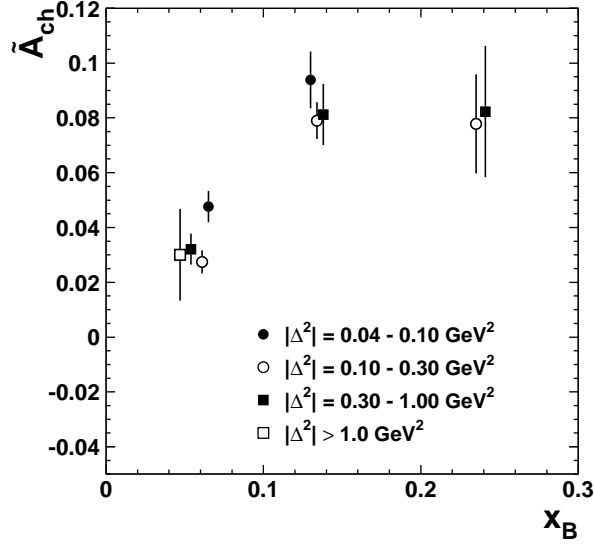,width=8.0cm}
\caption{\it Projected statistical accuracy for the $\phi$-integrated
lepton charge asymmetry Eq. {\rm(\ref{eq:asychx})} as a function of $x_B$ and 
$\Delta^2$ based upon GPD model version (E).}
\label{fig:lchxdep}
\end{figure}

Projections for the statistical accuracy attainable in measuring the lepton 
charge asymmetry are presented in fig. \ref{fig:lchphidep}. The results are
shown as a function of the azimuthal angle $\phi_\gamma$ in two regions of 
$x_B$, using different GPD parameterizations. The asymmetry clearly depends on
the particular parameterization and can even change its sign in dependence
on $x_b$. Therefore, in the experiment the asymmetry will have to be studied 
differentially as much as possible. Apparently, the inclusion of the $D$-term 
leads to essential changes in the asymmetry and future measurements at HERMES 
may allow to confirm its importance experimentally. In fig. \ref{fig:lchxdep} 
the integrated lepton charge asymmetry $\widetilde{A}_{ch}$ is shown as a 
function of $x_B$ and $\Delta^2$, based on the GPD parameterization (E). No
$\Delta^2$-dependence is seen on the basis of the chosen GPD model.

\begin{figure}[ht]
\centering
\begin{minipage}[c]{8.0cm}
\centering
\epsfig{file=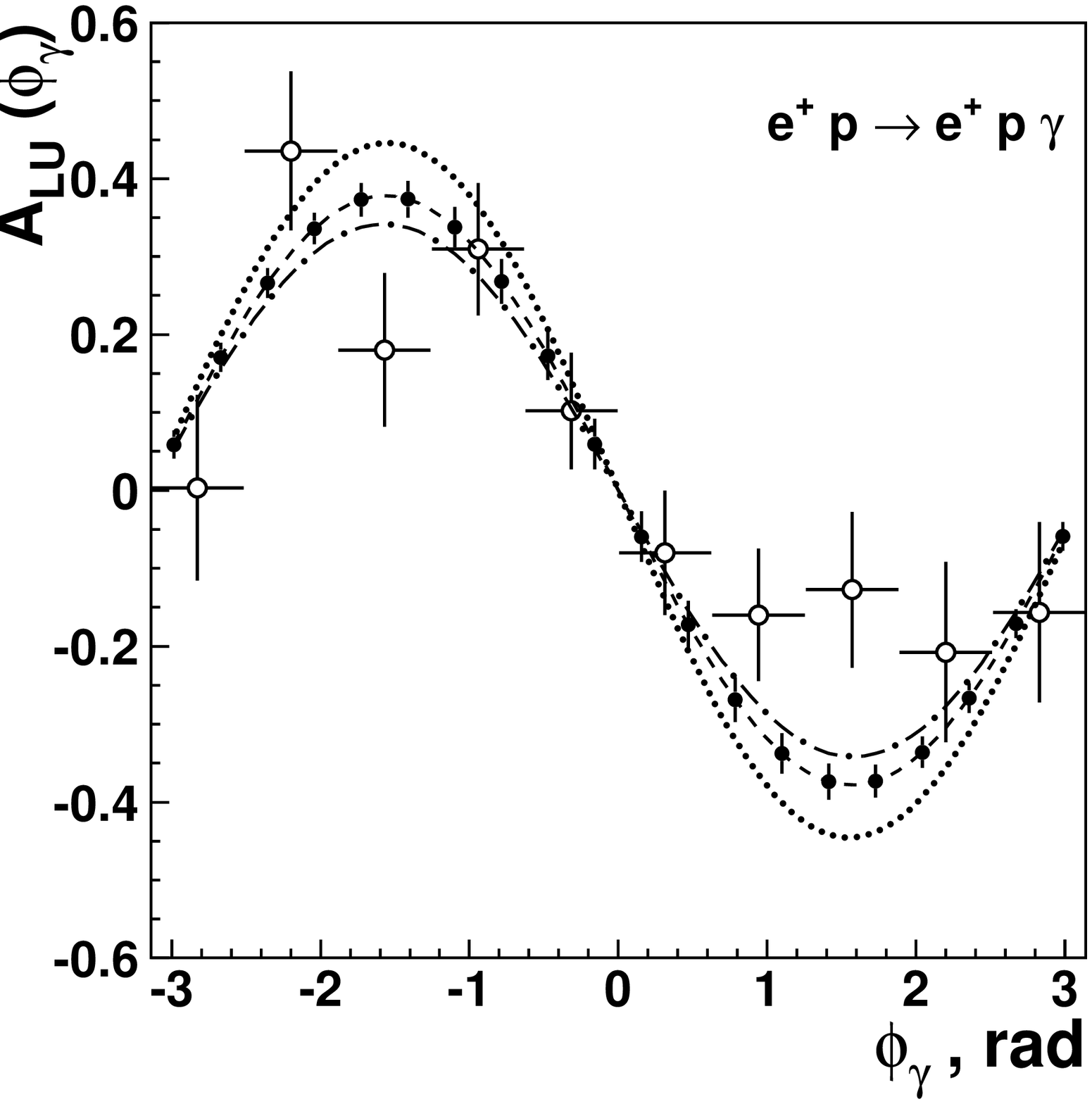,width=7.5cm}
\end{minipage}
\begin{minipage}[c]{8.0cm}
\centering
\epsfig{file=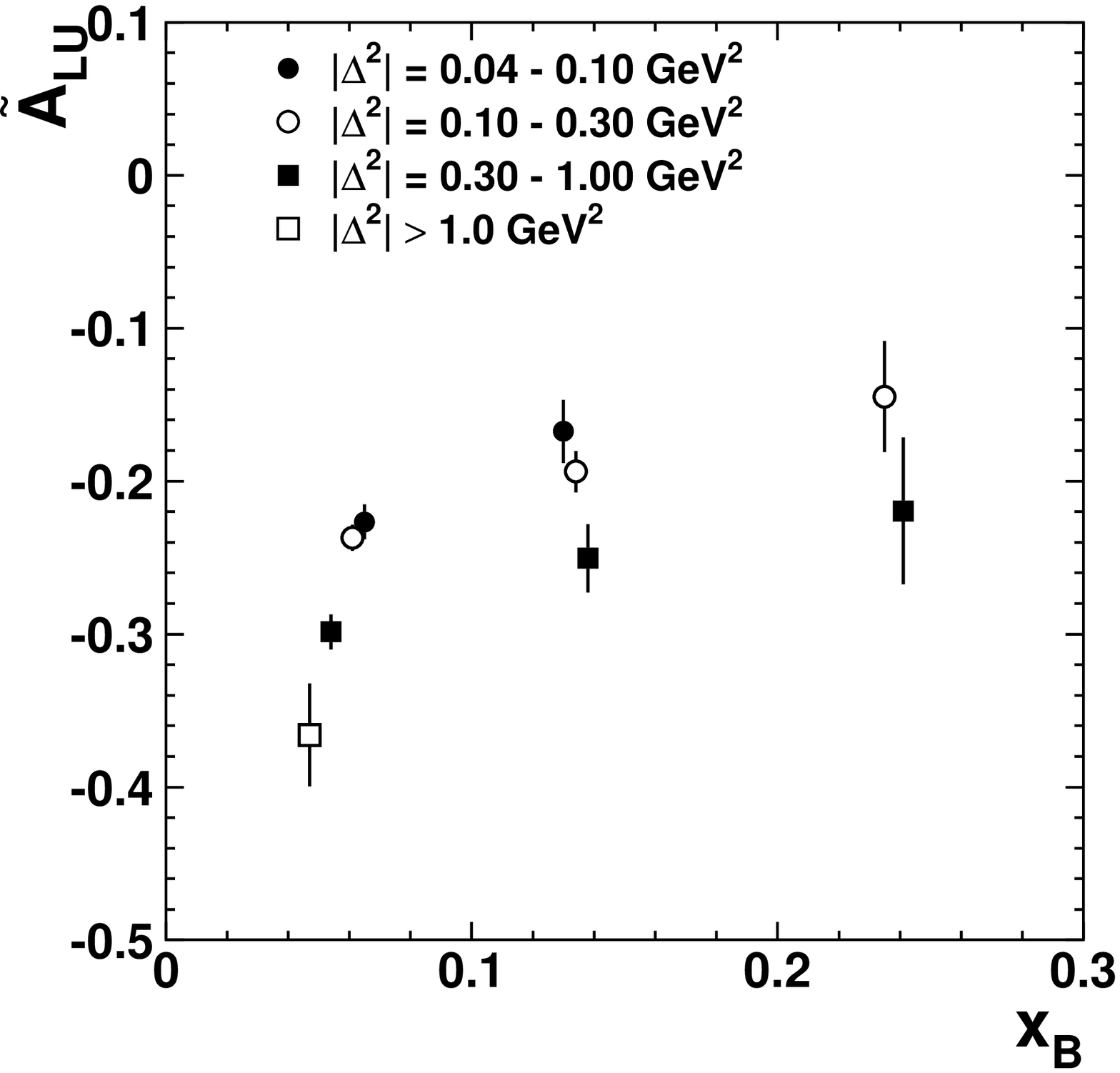,width=7.5cm}
\end{minipage}
\caption{\it Projected statistical accuracy for the beam helicity asymmetry 
Eq.s {\rm(\ref{eq:asybhphi}, \ref{eq:asybhx})} as a function
of the azimuthal angle $\phi_\gamma$ between scattering plane and production 
plane (left panel, closed circles) and as a function of $x_B$ and $\Delta^2$ 
(right panel). The line conventions are as in fig.\ref{fig:lchphidep}. The 
present measurement of the azimuthal asymmetry by HERMES is shown in the 
left panel by open circles.}
\label{fig:beamhel}
\end{figure}

Projections for the statistical accuracy attainable in measurements of the 
helicity asymmetry are presented in fig. \ref{fig:beamhel}, as a function
of the azimuthal angle $\phi_\gamma$ (left panel) and as a function 
of $x_B$ and $\Delta^2$ (right panel). As can be seen, the projections of 
the statistical accuracy promise a considerable improvement compared to 
the present HERMES measurement shown additionally in fig. \ref{fig:beamhel} 
(left panel). The projected asymmetry changes only slightly in dependence 
on the GPD parameterizations when the parameter $b$ of the profile function 
Eq. (\ref{eq:proffun}) varies in the range (1 ; $\infty$). Note that the 
beam helicity asymmetry is not sensitive to the D-term as it does not 
contribute to the imaginary parts of DVCS amplitudes.

\newpage

The considerations presented above are based upon the leading twist-2 level 
using the amplitudes calculated in Ref. \cite{belitsky}. More elaborate 
analyses include twist-3 effects \cite{kivpolvan,belitsky3} and 
next-to-leading order calculations \cite{freund}. Nevertheless, in the 
present situation where the generalized parton distributions
are practically unknown, the approach adopted in this paper appears
adequate to evaluate the expected size of the predictions. 

It appears worth noting that additional important constraints on GPDs can 
be expected from the analysis of the same data set by studying hard
exclusive production of both pseudo-scalar and vector mesons. 
As compared to DVCS these processes will provide information on different
combinations of generalized parton distributions.

\section{Conclusions}
Expected statistical accuracies have been evaluated for future measurements 
of Deeply Virtual Compton Scattering with the HERMES spectrometer
complemented by a to-be-built recoil detector. Using polarized electrons and
positrons of HERA with different helicities, in conjunction with an 
unpolarized proton target, it becomes possible to measure the lepton charge 
asymmetry and the lepton beam helicity asymmetry, both induced by the 
interference of the DVCS and the Bethe-Heitler process. From these 
asymmetries the real and the imaginary part of a certain DVCS amplitude 
combination can be determined.

The expected size of the asymmetries has been evaluated using various
parameterizations of the underlying generalized parton distributions. 
A number of different parameterizations has been used to 
compensate as much as possible for the present poor knowledge on GPDs. 
The level of the attainable statistical accuracy is mainly determined by the
cross section of the Bethe-Heitler process that dominates the reaction 
$e p \rightarrow e p \gamma$ at HERMES energies.

It has been shown that the planned measurements of hard exclusive real 
photon production at HERMES will be of high statistical significance. 
The envisaged separate results on the real and the imaginary part of a 
certain DVCS amplitude combination will constitute an important step 
towards the determination of the generalized parton distributions. 
The measured constraints will serve as very useful 
input for the further modelling of generalized parton distributions.

We thank R. Kaiser for careful reading of the manuscript.


\begin{thebibliography}{99}
\bibitem{muller94}
D.~M\"uller et al., Fortsch. Phys. {\bf 42}, 101 (1994).
\bibitem{Ji}
X.~Ji, Phys.Rev.Lett. {\bf 78}, 610 (1997); Phys.Rev. {\bf D55}, 7114 (1997).
\bibitem{rad}
A.V.~Radyushkin, Phys.Lett. {\bf B380}, 417 (1996); 
Phys.Rev. {\bf D56}, 5524 (1997).
\bibitem{radreview}
A.V.~Radyushkin, hep-ph/0101225.
\bibitem{hermesdvcs}
A.~Airapetian et al., hep-ex/0106068
\bibitem{CLASdvcs}
S.~Stepanyan et al., hep-ex/0107043
\bibitem{prcrecoil}
A Large Acceptance Recoil Detector for HERMES, DESY PRC 01-01, April 2001.
\bibitem{desyzeus}
P.R.B.~Saull, Proc. of the Int. Europhysics Conf. on HEP, Tampere, Finland,
15-21 July 1999, ed. by K.~Huitu, H.~Kurki-Suonio and J.~Maalampi 
(hep-ex/0003030); \\
L.~Favart, DIS-2001, Bologna, 27 April - 1 May (hep-ex/0106067).
\bibitem{desyh1}
C.~Adloff et al., DESY 01-093, 2001.
\bibitem{jisong}
X.~Ji, W.~Melnitchouk, X.~Song, Phys.Rev. {\bf D56}, 5511 (1997).
\bibitem{petrov98}
V.~Petrov et al., Phys.Rev. {\bf D57}, 4325 (1998).
\bibitem{vgg99}
M.~Vanderhaeghen, P.A.M.~Guichon, M.~Guidal, Phys.Rev. {\bf D60}, 
094017 (1999).
\bibitem{gpv01}
K.~Goeke, M.V.~Polyakov, M.~Vanderhaeghen, hep-ph/0106012.
\bibitem{polweiss}
M.V.~Polyakov, C.~Weiss, Phys.Rev. {\bf D60}, 114017 (1999).
\bibitem{kivpolvan}
N.~Kivel, M.V.~Polyakov, M.~Vanderhaeghen, Phys.Rev. {\bf D63}, 114014
(2001).
\bibitem{mrst98}
A.D.~Martin et al., Eur.~Phys.~J. {\bf C4}, 463 (1998).
\bibitem{leader}
E.~Leader, A.V.~Sidorov, D.B.~Stamenov, Phys.Rev. {\bf D58}, 114028 (1998).
\bibitem{belitsky}
A.V.~Belitsky et al., Nucl.Phys. {\bf B593}, 289 (2001).
\bibitem{belitsky3}
A.V.~Belitsky et al., Phys.Lett. {\bf B510}, 117 (2001). 
\bibitem{freund}
A.~Freund and M.~McDermott, hep-ph/0106124.
\end{thebibliography}
\end{document}